\documentclass[pdflatex,sn-mathphys-num]{sn-jnl}

\usepackage{amsmath}
\usepackage{amssymb}
\usepackage{graphicx}
\usepackage{multirow}
\usepackage{booktabs}
\usepackage{xcolor}
\usepackage{algorithm}
\usepackage{algpseudocode}
\usepackage{setspace}

\begin{document}

\doublespacing

\title{Well-Designed k-Space Coverage is Important for \protect\\[-2ex] Good MRI Denoising}

\author[1]{\fnm{Jiayang} \sur{Wang}}\email{jiayangw@usc.edu}
\author[1]{\fnm{Justin P.} \sur{Haldar}}\email{jhaldar@usc.edu}

\affil[1]{\orgdiv{Signal and Image Processing Institute, Ming Hsieh Department of Electrical and Computer Engineering}, \orgname{University of Southern California}, \orgaddress{\street{3740 McClintock Ave, EEB 400}, \city{Los Angeles}, \postcode{90089}, \state{CA}, \country{USA}}}

\abstract{\textbf{Object:} Modern computational MRI denoising approaches are often designed assuming fixed k-space coverage. This contrasts with earlier acquisition-design literature that leveraged k-space coverage modifications (e.g., reducing spatial resolution) to improve SNR. This work investigates whether the performance of modern computational denoising methods can be further enhanced by  k-space coverage modifications.

\textbf{Materials and Methods:} Using realistic simulations of noisy data, k-space coverage and averaging patterns were optimized for two advanced image denoising/reconstruction approaches: parallel imaging with total variation regularization and a U-Net neural network. For reference, comparisons against classical linear filtering/apodization methods were also performed. Performance was quantified using  normalized root-mean-squared error (NRMSE) and structural similarity (SSIM) metrics. 

\textbf{Results:} Advanced computational denoising methods can be substantially enhanced, both quantitatively and qualitatively, by reducing the spatial resolution of the acquisition to improve SNR. Indeed, even simple linear filtering/apodization with optimized k-space coverage can rival advanced methods using naive higher-resolution coverage.

\textbf{Discussion:} Classical acquisition design principles that allow spatial resolution to be traded for SNR enhancement are still very relevant for modern computational denoising techniques. However,  the optimization of k-space coverage and denoising/reconstruction methods can also be somewhat confounded because the NRMSE and SSIM metrics have low sensitivity to spatial resolution.}

\keywords{Image Enhancement; Image Reconstruction; Experimental Design}

\maketitle

\section{Introduction}\label{sec1}

Low SNR has always been one of the major limitations of MRI \cite{liang2000}. The field has invested decades of effort into mitigating this issue,  including advances in hardware, pulse sequences, and signal processing methods.  These efforts have broad potential significance because MRI is a flexible modality in which SNR is a tradable resource -- it can be sacrificed for benefits such as improved spatial resolution, faster acquisition, more informative image contrast, the ability to image species other than water protons, and reduced hardware costs.  As such, the ability to improve SNR has far-reaching implications for virtually all MRI applications, enabling more ambitious acquisitions that were previously impractical due to SNR limitations.

Recently,  there has  been substantial growth in the amount of research devoted to the computational ``denoising" of MRI data.  These approaches use advanced image reconstruction and/or post-processing methods to ``clean up" noisy images.  Importantly, although computational MRI denoising is not a new concept (see Ref.~\cite{haldar2022} for review of the early literature), the recent literature often approaches noise differently than  in older work.  In particular, recent literature often approaches MRI denoising problems in a compartmentalized way, with computational denoising steps kept fully isolated from data acquisition decisions.  It may not be coincidental that this compartmentalized approach closely resembles how generic image denoising problems are usually formulated in the modern image processing and computer vision literature.  A typical example is training a neural network  to accurately map a noisy image back to a corresponding ``noiseless" reference image using a formulation that is agnostic to acquisition details. 

While it can produce impressive results, this compartmentalized approach to denoising neglects the fact that denoising does not come for free.  Rather, the SNR improvement provided by denoising frequently comes -- implicitly or explicitly -- at the expense of  spatial resolution \cite{chan2021}.  Notably, computational denoising is also not the only mechanism for trading resolution for SNR in MRI. It has long been established that reducing spatial resolution  by shrinking the k-space coverage of the acquisition  can also very easily and dramatically improve SNR  \cite{edelstein1986,parker1990, macovski1996}.  Indeed, the earlier literature often put much more emphasis on this latter approach without much consideration of computational denoising.  In part, this is because denoising low-SNR data  was often viewed as inefficient in the SNR/resolution tradeoff relative to directly acquiring lower-resolution images  \cite{edelstein1986, haldar2022}.  This view was compounded by perceptions that high-resolution k-space samples carried little to no useful information when the SNR was poor \cite{watts2002}. It was also known theoretically that simple computational denoising methods like spatial averaging or linear spatial filtering/apodization  would have suboptimal SNR efficiency unless k-space averaging was specially tailored to match subsequent reconstruction/processing steps\cite{edelstein1986,parker1987,parker1990,mareci1991,ponder1994,hugg1996,star-lack1999,greiser2003,stobbe2008}.  

Despite these somewhat-negative early impressions, perceptions of computational denoising have improved considerably over time. It has more recently been theoretically demonstrated \cite{haldar2022},\cite{haldar2008},\cite{haldar2011},\cite[Sec.~3.3.2]{haldar2011c},\cite{haldar2012} that computational denoising methods can  be substantially more efficient in the SNR/resolution tradeoff than previously believed, and that high-resolution k-space samples can carry useful information even when the SNR is poor. There are also many recent publications, potentially numbering in the tens of thousands  \cite{milanfar2025}, demonstrating the empirical benefits of computational denoising.  Nonetheless, denoising performance is strongly influenced by the noise and resolution characteristics of the acquired data, and even good computational denoising approaches may be substantially limited by poor acquisition design.  Indeed,  Refs.~\cite{haldar2011,haldar2011c} found that, for one class of denoising approaches, modest noise reductions were efficient in the SNR/resolution tradeoff and cost only a small amount of spatial resolution.  In contrast, attempting very large noise reductions was inefficient relative to reducing the spatial resolution of the acquisition, suggesting that it may be preferable to avoid solving overly-challenging denoising problems when acquisition modifications could be used instead.   In this work, we investigate the extent to which such acquisition design principles for improving SNR can enhance the performance of modern computational denoising methods.

K-space sampling optimization has recently been receiving substantial  attention in the context of designing undersampling patterns for high-SNR MRI applications \cite{seeger2010,gozcu2018,haldar2019,bahadir2020,sherry2020,aggarwal2020,wang2022a}, building on foundations laid by earlier work \cite{vonkienlin1991,cao1993,marseille1994,reeves1995}.  However, similar recent interest has not been directed towards optimizing  k-space coverage or averaging patterns for low-SNR applications.   One exception is prior work demonstrating that the SNR efficiency of relatively simple (quadratic or quasi-quadratic) regularized reconstruction techniques can be improved using optimized k-space coverage and averaging \cite[Sec.~3.6]{haldar2011c}, but this result is not widely known and has not been extended to more advanced reconstruction techniques. 

 Our goal in this work is to revisit the effects of k-space optimization in the low-SNR regime with advanced computational denoising. Rather than attempting to find globally optimal sampling patterns  (which would require methodological innovation and draw focus away from our primary contribution), this work instead focuses on two relatively simple manipulations of MRI data acquisition with an aim to illuminate general and broadly-applicable  principles, closely following   classic analyses of SNR efficiency in MRI \cite{edelstein1986, parker1990}.  In the first case, we manipulate k-space coverage to reduce the resolution of a Nyquist-sampled acquisition, using any time savings to further improve SNR through additional data averaging.  In the second case, we build on the first case by additionally allowing nonuniform averaging of different parts of k-space.  Nonuniform averaging has been commonly advocated as a method for improving SNR efficiency in the earlier literature  \cite{parker1987,parker1990,mareci1991,ponder1994,hugg1996,star-lack1999,greiser2003,stobbe2008}.  Both cases are practically relevant and simple enough to be addressed using classical optimization techniques.

Our investigation is performed using two popular advanced image denoising/reconstruction approaches: parallel imaging reconstruction with total variation  regularization (SENSE-TV) \cite{block2007,lustig2007} and neural network reconstruction using a U-Net architecture \cite{zbontar2019}.  For reference, we also compare against much simpler linear filtering/apodization methods \cite{liang2000}.  For both SENSE-TV and U-Net reconstructions, our results demonstrate that substantial improvements in apparent denoising performance can be obtained when k-space coverage is optimized, and that further improvements are also possible with nonuniform averaging.  Indeed, the effects of optimized k-space sampling can be so dramatic that even simple linear filtering/apodization methods with optimized sampling can compete with naive SENSE-TV or U-Net reconstructions without sampling optimization.

A preliminary account of portions of this work was previously presented in \cite{wang2022}.

\section{Materials and Methods}\label{sec2}

\subsection{Theory}

The following subsections describe the specific noise modeling and k-space sampling assumptions we make for our investigation, as well as the methods we use to optimize k-space coverage and nonuniform averaging patterns.
\subsubsection{Noise Modeling Assumptions}
For concreteness and without loss of generality, we consider an acquisition based on 2D Cartesian k-space sampling with one phase encoding dimension  and one readout dimension, with a single readout  acquired per TR.  Notably, this choice does not limit the relevance of our results, since the mathematical relationship between spatial resolution and noise in Cartesian MRI  transcends these sequence-specific assumptions \cite{macovski1996, liang2000}.  As a result, even though the underlying mechanisms of noise behavior may vary,   mathematically equivalent resolution/noise tradeoffs would also be observed in other 2D Cartesian settings. For example, our simulations would be largely identical if we instead considered data acquired with two phase encoding directions, as frequently encountered in slice-by-slice reconstruction of 3D Cartesian imaging data (after transforming the fully-sampled readout), or if we  considered acquisitions that collect multiple readout lines per TR.

We consider an MRI experiment that acquires noisy k-space measurements on a centered Nyquist-rate Cartesian grid of k-space locations. For simplicity, we assume a square $N \times N$ sampling grid with even $N$, although our basic approach is easily generalized to arbitrary rectangular grids.  We further assume that data is measured from an array of $L$ receiver coils. This allows us to represent the noisy data using the model:
\begin{equation}
d_{\ell m n} = s_\ell[m,n] + z_{\ell m n},\label{eq:avgorig}
\end{equation}
for $m = -N/2,\ldots, N/2-1$, $n = -N/2,\ldots,N/2-1$, and $\ell=1,\ldots,L$.  In this expression, $s_\ell[m,n]$ is the ideal (noiseless) k-space sample corresponding to the $\ell$th channel, $m$th phase encoding position, and $n$th readout position, and $z_{\ell m n}$ represents the corresponding measurement noise.  

Following standard practice in multichannel MRI noise modeling, we model every noise sample  as complex-valued zero-mean circular Gaussian random noise \cite{roemer1990,pruessmann2001, robson2008}. We also assume that the noise in the multichannel data has been prewhitened \cite{pruessmann2001}, such that every noise sample is independent and identically distributed (i.i.d.), with no interchannel correlation. We use $\sigma^2$ to denote the common variance shared by each individual noise sample.  We also consider the scenario where the acquisition is able to acquire each phase-encoding line multiple times for the sake of SNR improvement through data averaging, and use $\tilde{w}_m$ to represent the integer number of measurements (averages) for the $m$th line.  After averaging, the noise $z_{\ell m n}$ will be complex-valued zero-mean circular Gaussian with (potentially $m$-dependent) variance $\sigma^2/\tilde{w}_m$.

\subsubsection{k-Space Sampling Assumptions}\label{sec:sampling}

As already mentioned, we focus on two potential manipulations of data acquisition, one involving reduced k-space coverage with uniform averaging (i.e., $\tilde{w}_m$ is the same for all $m$) and the other involving reduced k-space coverage with potentially non-uniform averaging (i.e., $\tilde{w}_m$ can be independently adjusted for each phase encoding line).   The former option is simpler and generally available by default on most MRI scanners (no pulse sequence programming required), while the latter option is more flexible and can yield higher SNR-efficiency \cite{parker1987,parker1990,mareci1991,ponder1994,hugg1996,star-lack1999,greiser2003,stobbe2008,haldar2011c} but generally requires pulse sequence modifications to enable full control over the nonuniform averaging pattern.

In both cases, we treat the acquisition gridsize $N$ as an optimization variable that we use to manipulate k-space coverage, with smaller $N$ resulting in lower spatial resolution.  We also  assume that we are given some upper limit $N_0$,  representing a practical limit on the nominal spatial resolution of the acquisition such that  $N \leq N_0$.   Notably, even without additional modifications to the acquisition, smaller values of $N$ are naturally associated with higher SNR \cite{liang2000}.   However, reducing $N$ without making other sequence changes can make comparisons difficult, since changing $N$ will also impact important experimental factors such as the total experiment duration and the duration of the readout.  To ensure fairness and avoid such confounds, it is common in the literature on SNR efficiency  (e.g., \cite{edelstein1986, macovski1996}) to fix both the total experiment duration and the duration of the readout, and we adopt the same approach herein.  This isolates the effects of altering the k-space coverage and averaging strategy from other experimental considerations, enabling well-controlled comparisons.

For convenience, we will describe the assumptions we make about modified k-space sampling patterns with reference to a hypothetical maximum-resolution MRI experiment that acquires data on an $N_0\times N_0$ sampling grid with uniform averaging (i.e., $\tilde{w}_m = \tilde{w}_0$ for all $m$, for some integer $\tilde{w}_0$).  The total acquisition time of this hypothetical experiment will be proportional to $\tilde{w}_0 N_0$, the total number of TRs needed to measure each of the $N_0$ phase encoding positions $\tilde{w}_0$ times.  Since uniform averaging is a special case of non-uniform averaging, our description below will focus on the non-uniform case.

Holding the total experiment duration constant means that if we reduce the value of $N$  below $N_0$, we must correspondingly change the number of averages $\tilde{w}_m$ for each phase encoding location such that the total number of TRs remains constant.  Intuitively speaking, if we have fewer phase encoding positions to measure, we will then have additional time that can be spent on increased averaging of the remaining phase encoding positions.  Mathematically, this corresponds to the constraint that
\begin{equation}
\sum_{m=-N/2}^{N/2-1} \tilde{w}_m = \tilde{w}_0 N_0.\label{eq:const}
\end{equation} 

In addition, maintaining the readout duration when the number of readout samples $N$ is reduced below $N_0$ often involves changes  to the readout gradient amplitude and the acquisition bandwidth. These changes are beneficial since they lead to SNR-improvements akin to averaging \cite{macovski1996}.  Specifically, if the variance of the unaveraged noise had originally been $\sigma^2$ for the hypothetical high-resolution reference experiment, then after making appropriate bandwidth adjustments to accommodate $N$ readout samples with the same readout duration, the variance of the averaged noise $z_{\ell mn}$ in the modified acquisition will become $\sigma^2/(\tilde{w}_m N_0/N)$.  

For notational simplicity, we define an ``effective" number of averages $w_m$ to combine the effects of phase-encode averaging and readout bandwidth adjustments, with $w_m = \tilde{w}_m N_0/N$.  With this choice, the noisy averaged data samples $d_{\ell mn}$ can still be modeled as in  Eq.~\eqref{eq:avgorig}, except that the variance of $z_{\ell mn}$ is now given by $\sigma^2/w_{m}$. The previous constraint on the total scan time from Eq.~\eqref{eq:const} can also be rewritten in terms of the effective number of averages as
\begin{equation}
\sum_{m=-N/2}^{N/2-1} w_m  = \tilde{w}_0 N_0^2/N.
\end{equation}

\subsubsection{k-Space Sampling Optimization}\label{sec:opt}
We formulate our  sampling optimization problem in a generic way that is compatible with arbitrary image reconstruction techniques, with optimization performed with respect to a database of high-quality training data.   Specifically, we perform optimization of both the k-space acquisition (i.e., the gridsize and effective number of averages)  and the reconstruction parameters using a loss function that quantifies the average difference between the denoised/reconstructed image and the corresponding reference image across the training set. 

To enable easy quantitative performance comparisons between images produced from different acquisition gridsizes $N$, we assume that images are consistently reconstructed on an $N_0\times N_0$ grid regardless of the value of $N$.  Specifically, we assume a generic reconstruction function of the form
\begin{equation}
\hat{\mathbf{x}} = g_{\mathbf{p},N,\mathbf{w}}( \mathbf{d}),\label{eq:refun}
\end{equation}
where $\hat{\mathbf{x}} \in \mathbb{C}^{N_0^2}$ is the $N_0 \times N_0$ reconstructed image, $\mathbf{d} \in \mathbb{C}^{N^2L}$ is the vector of averaged data samples $d_{\ell m n}$, $\mathbf{w} \in \mathbb{C}^N$ is the vector of the effective number of averages $w_m$ for each phase encoding position, and $\mathbf{p}$ is the  vector of trainable parameters of the reconstruction method (e.g., the regularization parameter for SENSE-TV reconstruction, apodization weights for simple linear filtering, or neural network weights for U-Net reconstruction).  

For training, we assume access to a set of $T$ ``noise-free" high-resolution k-space datasets $\mathbf{s}_t \in \mathbb{C}^{N_0^2 L}$ for $t=1,\ldots,T$.  The entries of each $\mathbf{s}_t$ vector correspond to dataset-specific instances of the ideal samples $s_\ell[m,n]$ from Eq.~\eqref{eq:avgorig} over an $N_0 \times N_0$ grid.  For each candidate acquisition gridsize $N$, we extract a lower-resolution version $\mathbf{s}_t^N \in \mathbb{C}^{N^2L}$ by discarding samples that would fall outside an  $N\times N$ grid.  We also form a corresponding low-resolution ``noise-free" reference image $\mathbf{r}_t^N \in \mathbb{C}^{N_0^2}$ by zero-padding $\mathbf{s}_t^N$ to an $N_0\times N_0$ grid, applying an inverse Fourier transform, and combining channels using known coil sensitivity maps \cite{pruessmann1999}.  Our decision to use low-resolution reference images instead of high-resolution reference images was based on the behavior we observed for  U-Net reconstruction, whose performance degraded if high-resolution reference images were used for training.  We specifically observed substantial hallucination of high-frequency content in this scenario --  likely because this setup resembles a super-resolution task, which is often substantially more difficult than denoising  \cite{haldar2022}.  

To obtain realistic simulations of noisy measurements that are consistent with our noise model (i.e., Eq.~\eqref{eq:avgorig} with the updated variance described in Sec.~\ref{sec:sampling}), we start by generating pseudo-random noise vectors $\mathbf{z}_t^N \in \mathbb{C}^{N^2L}$ with zero-mean i.i.d. circular complex Gaussian entries of variance $\sigma^2$. For a given $N$ and $\mathbf{w}$, this allows us to produce simulated noisy data with realistic noise statistics using;
\begin{equation}
\mathbf{d}_t^N = \mathbf{s}_t^N + \frac{1}{\sqrt{\mathbf{w}}} \odot \mathbf{z}_t^N,
\end{equation}
where $\odot$ represents elementwise multiplication, and through a slight abuse of notation, $\frac{1}{\sqrt{\mathbf{w}}}$  represents the length-$N^2 L$  vector obtained by setting all of the entries entries corresponding to the $m$th phase encoding position (across all readout points and channels) equal to $1/\sqrt{w_m}$. 

Based on this setup, we formulate the following constrained optimization problem to jointly optimize the acquisition parameters $N$ and $\mathbf{w}$ and reconstruction parameters $\mathbf{p}$:
\begin{equation}
\begin{split}
\hat{N}& = \arg\min_{N}  \sum_{t=1}^T J\left(g_{\hat{\mathbf{p}}(N),N,\hat{\mathbf{w}}(N)}\left(\mathbf{s}_t^N+ \frac{1}{\sqrt{\hat{\mathbf{w}}(N)}} \odot \mathbf{z}_t^N\right), \mathbf{r}_t^{N_0} \right) 
\end{split}
\label{eq:loss}
\end{equation}
subject to
\begin{equation}
\{\hat{\mathbf{p}}(N), \hat{\mathbf{w}}(N)\} = \arg\min_{\mathbf{p},\mathbf{w}}  \sum_{t=1}^T J\left(g_{\mathbf{p},N,\mathbf{w}}\left(\mathbf{s}^N_t+ \frac{1}{\sqrt{\mathbf{w}}} \odot \mathbf{z}_t^N\right), \mathbf{r}^N_t \right) \text{ subject to }  \|\mathbf{w}\|_1 = \tilde{w}_0 N_0^2/N.\label{eq:loss2}
\end{equation}
In this expression, $J(\mathbf{x},\mathbf{r})$ is a distance metric to quantify the discrepancy between the reconstructed image and the reference image.   Note that we have adopted a two-level optimization structure in which the reconstruction parameters $\hat{\mathbf{p}}(N)$ and the averaging pattern $\hat{\mathbf{w}}(N)$ is optimized with respect to low-resolution reference images for each $N$, while $N$ is chosen with comparison against the high-resolution reference image. This structure ensures that  $\hat{\mathbf{p}}(N)$ and $\hat{\mathbf{w}}(N)$ are obtained from solving a pure denoising problem, while still ultimately prioritizing high-resolution results when selecting $N$.

For optimization, we treat the reconstruction parameters $\mathbf{p}$ as  continuous variables that can be optimized using standard techniques.  For $\mathbf{w}$,  the true number of averages must ultimately be integer-valued.  While it is common in the literature on sampling design for undersampled MRI to employ advanced algorithms that directly enforce this constraint \cite{seeger2010,gozcu2018,haldar2019,bahadir2020,sherry2020,aggarwal2020,wang2022a, vonkienlin1991,cao1993,marseille1994,reeves1995}, such methods introduce complexity that is unnecessary for our purposes.  Instead,  following classical practice in  statistical experiment design \cite{federov1972, pukelsheim1993} and prior MRI literature on optimal averaging  \cite{mareci1991, parker1987}, we relax the integer constraint and allow the entries of $\mathbf{w}$ to take arbitrary positive real values.  This relaxation enables the use of standard  optimization techniques such as stochastic gradient descent and can sometimes even accommodate analytic solutions (see, e.g.,  \cite[Thm.~3.1]{haldar2011c} and related discussion), which is especially advantageous given that integer programming is generally NP-hard.  Importantly, this approach -- sometimes known as ``continuous design," ``approximate design," or ``design for infinite sample size" \cite{federov1972, pukelsheim1993} -- produces continuous solutions. These solutions can be converted into valid integer-valued solutions through an appropriate rounding procedure, without requiring that the desired total number of measurements $\tilde{w}_0N_0$ is fixed in advance.

The parameter $N$ should also be integer-valued, but unlike $\mathbf{w}$, a continuous relaxation is not straightforward or particularly useful in this case.  In this work, we use a  brute-force grid search over the finite set of  candidate $N$ values. Brute-force search is a classical and widely used approach in discrete optimization -- it is simple to implement, helps avoid certain types of local minima, and can be very efficient when the search space is small.

Taken together, this results in two different algorithms, depending on whether we assume uniform averaging or allow nonuniform averaging.  The algorithm for uniform averaging is:
\begin{itemize}
\item For each candidate value of $N$:
\begin{itemize}
\item Obtain $\hat{\mathbf{w}}(N)$ by setting $\hat{w}_m(N) = \tilde{w}_0 N_0^2/N^2$ for all $m$. (The averaging pattern is constant, and will not be treated as an optimization variable)
\item Iteratively optimize the reconstruction parameters $\hat{\mathbf{p}}(N)$ by solving Eq.~\eqref{eq:loss2} with fixed $\hat{\mathbf{w}}(N)$, using standard continuous optimization techniques (in this work, we use either stochastic gradient descent or the Adam optimizer, as described in the sequel).
\end{itemize}
\item Return the $\hat{N}$, $\hat{\mathbf{p}}(N)$, and $\hat{\mathbf{w}}(N)$ values that achieved the lowest value of the loss function  from Eq.~\eqref{eq:loss}.
\end{itemize}

The algorithm for nonuniform averaging is:
\begin{itemize}
\item For each candidate value of $N$:
\begin{itemize}
\item Initialize $\hat{\mathbf{w}}(N)$ by setting $\hat{w}_m(N) = \tilde{w}_0 N_0^2/N^2$ for all $m$.
\item Solve  Eq.~\eqref{eq:loss2} to obtain $\hat{\mathbf{p}}(N)$ and $\hat{\mathbf{w}}(N)$  using standard continuous optimization techniques as above. We use projected stochastic gradient descent to enforce the constraints on $\mathbf{w}$.  The optimal projection onto the constraint set can be derived using simple Lagrange multiplier methods, and is achieved by replacing each $w_m$ with $w_m+ \beta/N$, with $\beta = \tilde{w}_0 N_0^2/N -  \sum_{m=-N/2}^{N/2-1} w_m$.    
\end{itemize}
\item Return the $\hat{N}$, $\hat{\mathbf{p}}(N)$, and $\hat{\mathbf{w}}(N)$ values that achieved the lowest value of the loss function from Eq.~\eqref{eq:loss}.
\end{itemize}

After obtaining a continuous averaging design in terms of effective averages $\hat{\mathbf{w}}$, we convert back to actual averages and round to integer values using a simple (potentially suboptimal) rounding procedure.   Specifically, the final integer-valued averaging pattern is obtained by solving
\begin{equation}
\hat{\mathbf{q}} = \arg\min_{\mathbf{q} \in \mathbb{Z}^N} \| \mathbf{q} - N/N_0\hat{\mathbf{w}}\|_1 \text{ s.t. } \|\mathbf{q}\|_1 = \tilde{w}_0 N_0\label{eq:cost}
\end{equation}
An optimal solution to this problem can be obtained by first rounding each entry of $N/N_0\hat{\mathbf{w}}$ to the nearest integer, i.e., $\hat{\mathbf{q}} = \mathrm{round}(N/N_0\hat{\mathbf{w}})$.  If the resulting  pattern $\hat{\mathbf{q}}$ sums to $\tilde{w}_0 N_0$, the process terminates because the rounded solution is optimal with respect to Eq.~\eqref{eq:cost}.  Otherwise, if $\hat{\mathbf{q}}$ includes too many or too few averages, we iteratively subtract or add one average at a time in a greedy manner until the sum constraint is satisfied. At each step,  one average is added or removed from the phase encoding position that will result in the smallest increase of the cost function Eq.~\eqref{eq:cost}.  In case of ties, averages nearest to the center of k-space are preferentially preserved.  It can be shown that, due to the properties of the $\ell_1$-norm, this greedy algorithm will produce a globally optimal solution to Eq.~\eqref{eq:cost}.

\subsection{Methods}\label{sec3}

\subsubsection{Data}
Our experiments are based on real T1-weighted brain MRI data from fastMRI  \cite{zbontar2019}.  Specifically, we used the central 8  slices of multichannel k-space data from 373 subjects (2984 slices in total).  These datasets were provided in a form representing an intermediate step in the standard oversampled analog-to-digital conversion process \cite[Ch.~4.8.1]{oppenheim1999}.  Specifically,  the data is $2\times$ oversampled after being filtered with an analog antialiasing filter, but the remaining digital filtering and downsampling steps have not yet been applied, which results in noise that is correlated along the readout dimension.  We completed the digital filtering and downsampling steps to ensure that our i.i.d. noise modeling assumptions would be valid.  After processing, each k-space dataset had a matrix size of $N_0\times N_0 = 320\times320$ (corresponding to a nominal in-plane resolution of 0.5mm$\times$0.5mm), with $L=16$ channels.

Notably, since these are real datasets, they are not truly ``noise-free", even though we will use them as the ``noise-free" high-resolution k-space datasets $\mathbf{s}_t$ in our formulation.  This is a potential concern, since it has been previously shown that the presence of ``hidden noise" in data that is assumed to be noise-free can sometimes cause interpretation problems \cite{wang2024}. However, we do not expect this to be a major confound for this study, since our simulated noise is substantially larger than the ``hidden noise" present in the data, and because we are also using sensitivity-based coil combination to generate reference images, which can help mitigate the effects of hidden noise \cite{wang2024}.

To facilitate later numerical optimization steps, each slice was independently normalized by dividing all its multichannel k-space samples by its corresponding maximum voxel magnitude  across all channels -- this ensures that the complex-valued multichannel images will each have voxel magnitudes in the range of $[0,1]$. The 2984 slices were partitioned into three sets: 2648 for training, 168 for validation (used for choosing parameters such as network architectures, learning rates, and number of training epochs), and 168 for testing.  Sensitivity maps for the SENSE model were obtained using PISCO\cite{lobos2023}.

\begin{figure}[t!]
    \centering
    \includegraphics[width=1\textwidth]{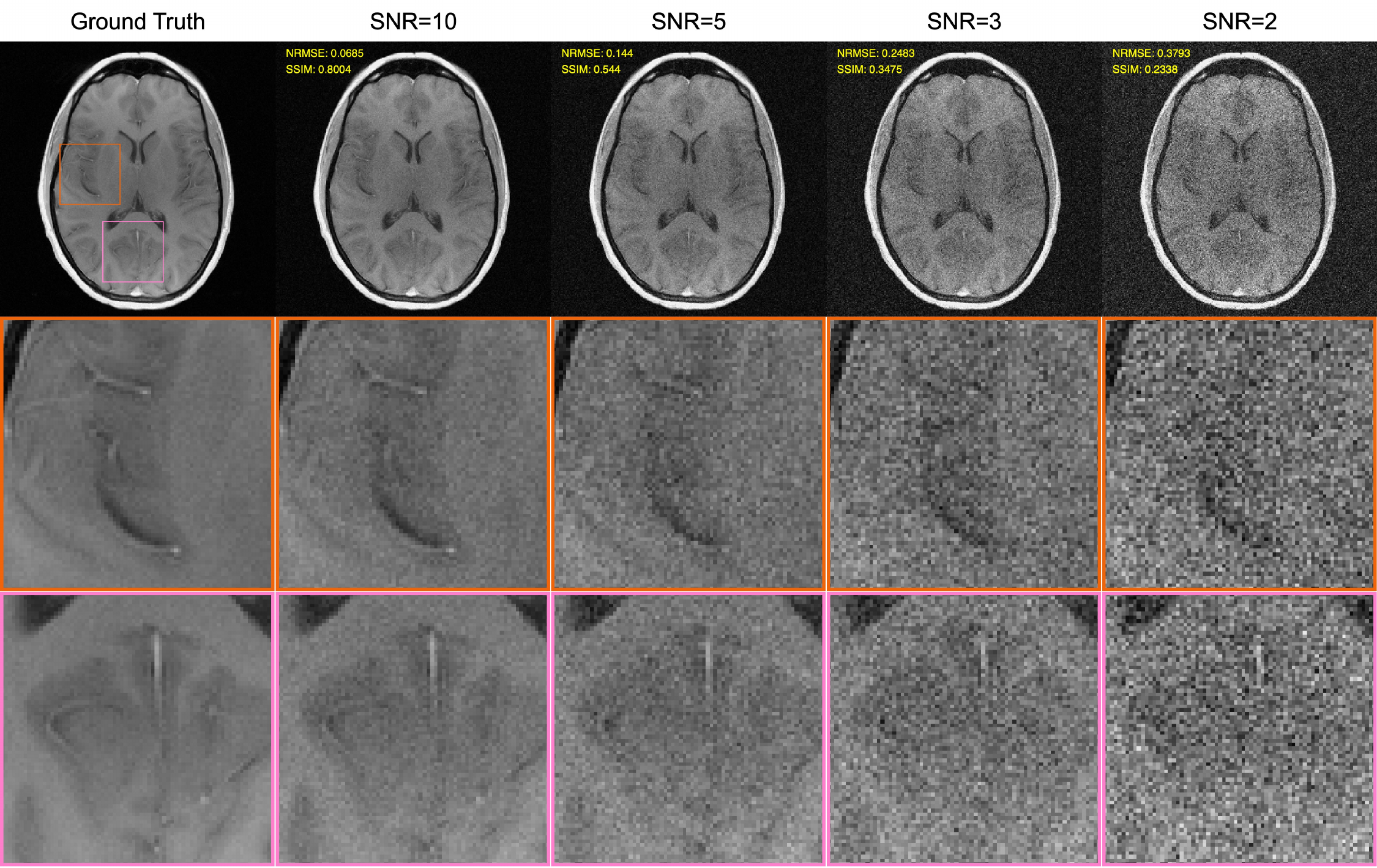}
\caption{Illustrative examples of the different SNRs we considered, shown for a representative slice. The top row shows the ground truth image $\mathbf{r}_t^{N_0}$, as well as simulated noisy images that are obtained for high-resolution ($N_0\times N_0$) sampling  with uniform averaging $\tilde{w}_0 = 8$ for different simulated noise levels. The images were obtained using simple inverse Fourier transform reconstruction of the averaged noisy data, followed by  SENSE-based coil combination.  We show normalized root-mean-squared error (NRMSE) and structural similarity (SSIM) metrics in the top left corner of each noisy image. The bottom two rows show zoom-ins to specific patches, with the patch locations marked in the ground truth image of the top row. }
\label{fig:SNR}
\end{figure}

Our simulations explored four different noise levels: SNR = 2, 3, 5, and 10, representing denoising problems with varying degrees of difficulty as illustrated in Fig.~\ref{fig:SNR}. The target SNR levels were defined for images generated using conventional reconstruction -- i.e., Fourier reconstruction with sensitivity-based coil combination.  The simulated noise variance was selected to produce the desired SNR levels in the central white matter of typical brain slices, assuming a high-resolution acquisition ($N_0$=320) with uniform 8$\times$ averaging (i.e., $\tilde{w}_0 = 8$).

\subsubsection{Training and Evaluation}
For each SNR level and each reconstruction method, the k-space sampling patterns and reconstruction parameters  were jointly optimized to minimize an objective function in the form of Eq.~\eqref{eq:loss}, as already described in Sec.~\ref{sec:opt}.  In all cases, training was performed using the mean-squared error loss function $J(\mathbf{x},\mathbf{r}) = \| \mathbf{x} - \mathbf{r}\|_2^2$. Practical constraints prevent us from comprehensively evaluating other choices of $J(\cdot,\cdot)$, although we anecdotally observed that other loss functions such as the $\ell_1$-norm or structural similarity (SSIM) \cite{wang2004b} produced similar optimized k-space averaging patterns. 

Results were evaluated using the 168 datasets that were set aside for testing.  Evaluations were based on comparing the denoising results against the corresponding reference images, including qualitative visual assessments as well as quantitative comparisons  using common measures such as the normalized root-mean-squared error (NRMSE), defined as $\|\mathbf{x}-\mathbf{r}\|_2/\|\mathbf{r}\|_2$), and SSIM \cite{wang2004b}.

Additional reconstruction-specific details are described below:

\noindent \underline{\bf SENSE-TV Denoising/Reconstruction}\\[2pt]
In this case, reconstruction was performed using Eq.~\eqref{eq:refun}, using the denoising/reconstruction function
\begin{equation}
g_{\mathbf{p},N,\mathbf{w}}( \mathbf{d}) = \arg\min_{\mathbf{x} \in \mathbb{C}^{{N_0}^2}} \|  \sqrt{\mathbf{w}} \odot ( \mathbf{E} \mathbf{x} - \mathbf{d}) \|_2^2 + \lambda \|\mathbf{D}\mathbf{x}\|_1.\label{eq:sensetv}
\end{equation}
Here, $\mathbf{E}$ is the forward model of data acquisition (representing the combined effects of sensitivity encoding by the multichannel receiver array, Fourier encoding, and low-resolution sampling on the $N\times N$ subset of the nominal $N_0 \times N_0$ k-space grid), $\mathbf{D}$ is a spatial finite difference operator, $\lambda$ is the TV regularization parameter, and $\sqrt{\mathbf{w}}$ appears in the data consistency term to properly model the known statistical characteristics of the averaged data $\mathbf{d}$. 

The solution to Eq.~\eqref{eq:sensetv} was obtained using an ADMM approach \cite[Sec.~6.4.1]{boyd2011} with the number of iterations fixed at 50.  For each noise level and each $N$, the ADMM penalty parameter -- an algorithmic parameter that does not appear in Eq.~\eqref{eq:sensetv} -- was roughly tuned to maximize NRMSE for a representative slice, after which it was held fixed. 

The optimizable reconstruction parameters $\mathbf{p}$  consisted solely of the TV regularization parameter $\lambda$.  In the uniform-averaging scenario, $\mathbf{p}$ was optimized using stochastic gradient descent (batch size = 8) with a learning rate of 0.01 and a decay factor of 0.9 per epoch for 10 epochs (further epochs yielded consistent results).  In the nonuniform-averaging scenario, $\mathbf{p}$ and $\mathbf{w}$ were jointly optimized using projected stochastic gradient descent (batch size = 8) to enforce the constraints on $\mathbf{w}$. Additionally, at each step, the gradients for $\mathbf{w}$ were normalized to have unit $\ell_2$-norm, which we found to be helpful for stabilizing the optimization process across different SNRs and reconstruction methods.  During the first 45 epochs, $\mathbf{w}$ and $\mathbf{p}$ were jointly optimmized using a  learning rate of 0.001 and a decay factor of 0.99 per epoch for $\mathbf{p}$, and a learning rate of 0.01 and a decay factor of 0.99 per epoch for $\mathbf{w}$.  Subsequently, the averaging pattern was rounded to integer values and fixed, and $\mathbf{p}$ was further optimized for an additional 5 epochs.

\noindent \underline{\bf U-Net Denoising/Reconstruction}\\[2pt]
In this case, denoising/reconstruction was performed using the U-Net implementation from Ref.~\cite{zbontar2019}, with a base number of 64 channels, 3 pooling layers, and no dropout.  The number of input channels was set to 32 (the  real and imaginary parts of the images obtained from applying the Fourier transform to the zero-filled k-space data from each of the $L=16$ channels), and the number of output channels was set to 2 (the real and imaginary parts of the coil-combined image).  

In the uniform-averaging scenario, training was performed using a batch size of 8, using the Adam optimizer for the neural network parameters $\mathbf{p}$ with a learning rate of 0.0003 for 100 epochs.    In the nonuniform-averaging scenario, $\mathbf{p}$ and $\mathbf{w}$ were optimized jointly (batch size = 8) for 90 epochs. The Adam optimizer  with a learning rate of 0.0003 was used for $\mathbf{p}$.  Projected stochastic gradient descent for  with a learning rate of 0.01 and a decay factor of 0.99 per epoch was used for $\mathbf{w}$, using gradient normalization as described previously.  Subsequently, the averaging pattern was rounded to integer values and fixed, and $\mathbf{p}$ was further optimized for an additional 10 epochs.

\noindent \underline{\bf Apodized Denoising/Reconstruction}\\[2pt]
For comparison against advanced methods like SENSE-TV and U-Net denoising/reconstruction, we also implemented simple linear spatial filtering/apodization techniques, which are some of the earliest data processing methods to enhance the SNR of MRI images  \cite{liang2000}, and for which optimal averaging patterns can be derived analytically \cite{parker1987,mareci1991}.  These methods multiply the measured data by a window function $h(k_{pe}, k_{ro})$ (e.g., a Hamming window or a Gaussian window) prior to simple Fourier reconstruction, such that the estimated image for each channel is obtained as
\begin{equation}
\hat{f}_\ell(x,y) = \Delta k_{pe} \Delta k_{ro}\sum_{m=1}^N \sum_{n=1}^N d_{\ell m n} h( m \Delta k_{pe}, n \Delta k_{ro}) e^{i2\pi (m \Delta k_{pe} x + n \Delta k_{ro} y)}.
\end{equation}
In the spatial domain, this has the effect of convolving the image obtained from conventional Fourier reconstruction with the spatial filter associated with the k-space window.  After this, the images for each-channel can by coil-combined using SENSE as described above to generate the image estimate $\hat{\mathbf{x}}$.  

In our implementation, rather than using a fixed predetermined window, we allow the window function to be optimized separately for each SNR in a data-driven way.  Specifically, the reconstruction parameters $\mathbf{p}$ consist of the $N^2$ window values $h( m \Delta k_{pe}, n \Delta k_{ro})$ for $m,n=-N/2,\ldots,N/2-1$, which we constrained to be real-valued and positive.  

In the uniform-averaging scenario, $\mathbf{p}$ was optimized using projected stochastic  gradient descent (batch size = 8) with a learning rate of 1 and a decay factor of 0.99 per epoch for 50 epochs. In the nonuniform-averaging scenario, $\mathbf{p}$ and $\mathbf{w}$ were jointly optimized using projected stochastic gradient descent to enforce the constraints on $\mathbf{p}$ and $\mathbf{w}$ (batch size = 8). A learning rate of 1 and a decay factor of 0.99 per epoch was used for $\mathbf{p}$. A learning rate of 0.1 and a decay factor of 0.99 per epoch was used for $\mathbf{w}$, using gradient normalization  as described previously.  After 45 epochs, the averaging pattern was rounded to integer values and fixed, and $\mathbf{p}$ was further optimized for an additional 5 epochs.

\section{Results}\label{sec4}

\begin{figure}[t]
    \centering
    \includegraphics[width=1\textwidth]{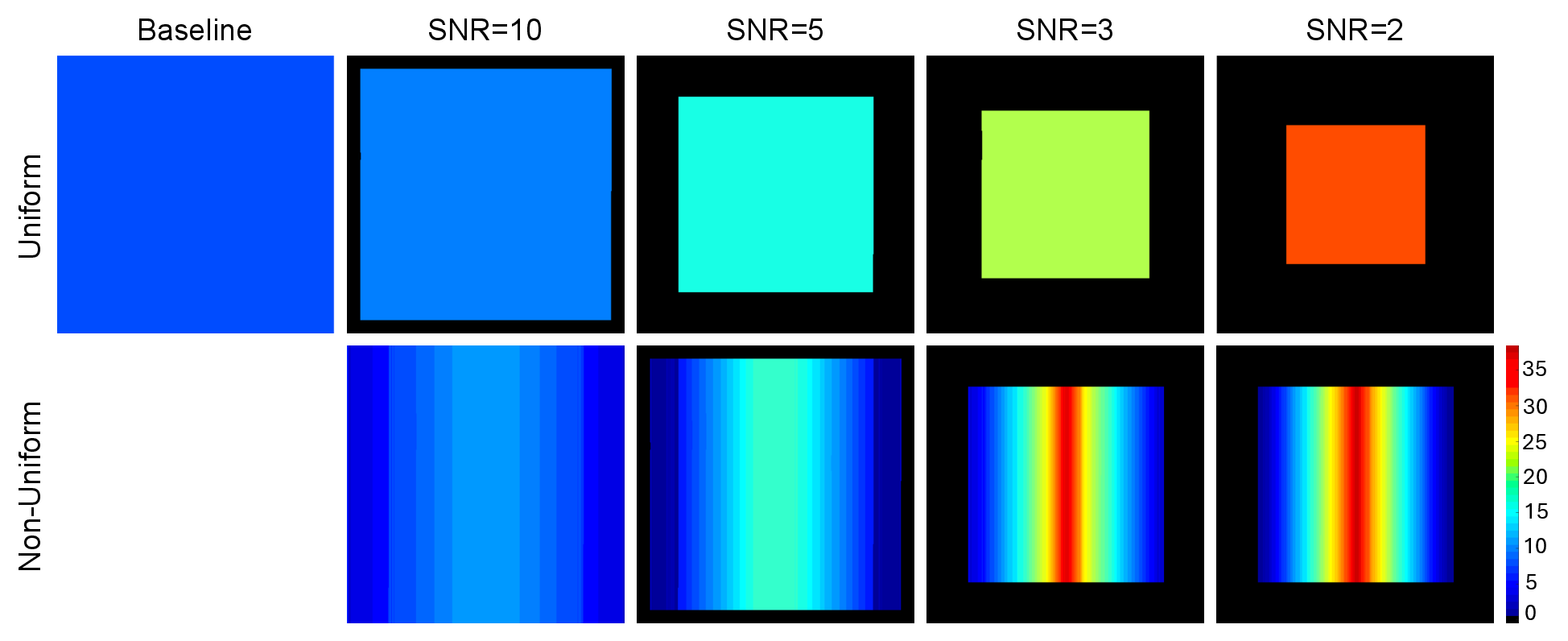}
\caption{Optimized averaging patterns corresponding to SENSE-TV denoising/reconstruction, obtained using (top row) uniform averaging and (bottom row) nonuniform averaging. For reference, the baseline acquisition approach ($8\times$ averaging of high-resolution $N_0\times N_0$ data) is shown in the top left. }
\label{fig:avg_stv}
\end{figure}

\begin{figure}[t]
    \centering
    \includegraphics[width=1\textwidth]{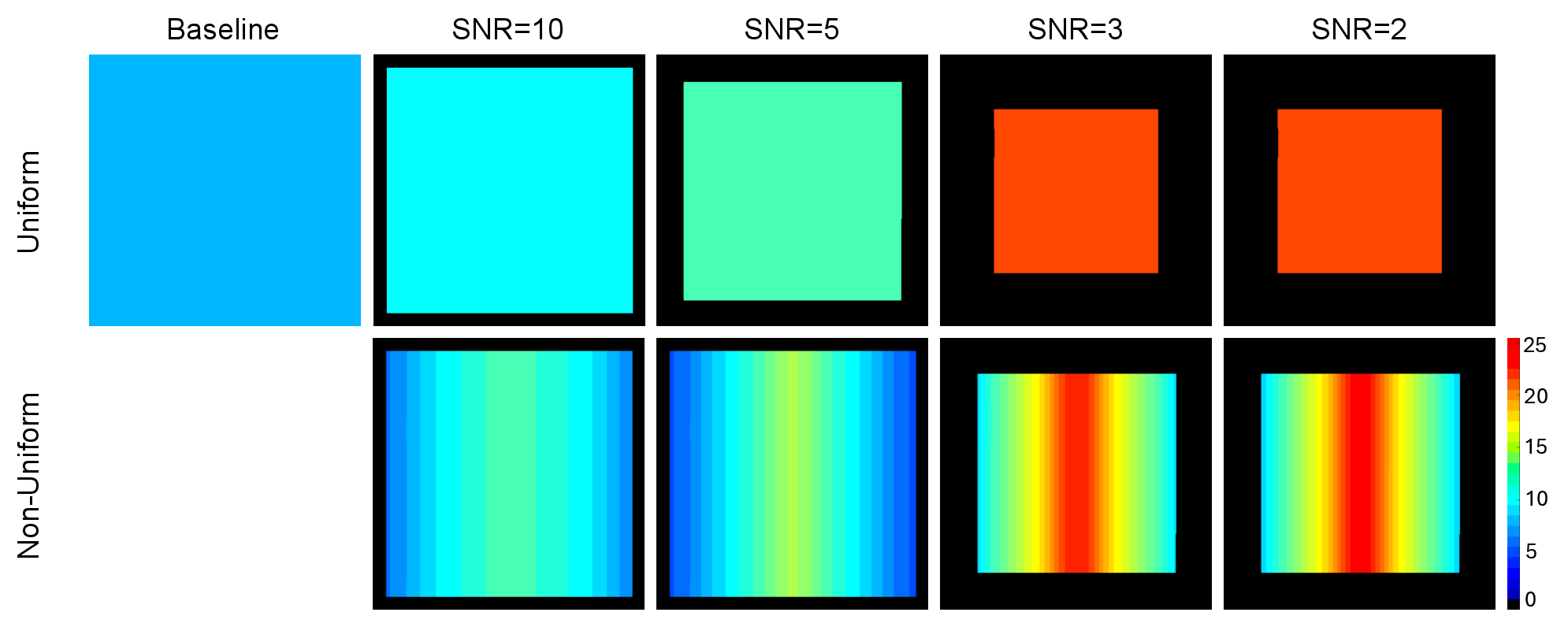}
\caption{Optimized averaging patterns corresponding to U-Net denoising/reconstruction, obtained using (top row) uniform averaging and (bottom row) nonuniform averaging. For reference, the baseline acquisition approach ($8\times$ averaging of high-resolution $N_0\times N_0$ data) is shown in the top left. }
\label{fig:avg_unet}
\end{figure}

The optimized averaging patterns we obtained for SENSE-TV and U-Net denoising/reconstruction are shown in Figs.~\ref{fig:avg_stv} and \ref{fig:avg_unet}, respectively.  As can be seen, we observed in the uniform-averaging scenario that lower-resolution acquisitions ($\hat{N} \ll N_0$) were increasingly favored over higher-resolution acquisitions as the SNR grew lower. This is different from the SNR-agnostic acquisition strategy that is typically employed in studies of computational MRI denoising, although is well-aligned with the classical MRI experiment design strategy of employing reduced resolution to mitigate poor SNR.  

In the nonuniform averaging scenario, we again observed that smaller $\hat{N}$ values were favored as the SNR grew lower. In addition,  we also observed that the averaging pattern shifted to prefer more averaging of low-frequency k-space than of higher-frequency k-space.  Interestingly, these optimized averaging patterns for SENSE-TV and U-Net reconstruction have similar characteristics to the optimized averaging patterns that have been previously derived for simple linear filtering/apodization methods \cite{parker1987,mareci1991} or for simple quadratic regularized reconstruction methods \cite[Sec.~3.6]{haldar2011c}, which also devoted more effort to averaging low-frequency k-space.  However, it is also notable that the relative proportion of low-frequency versus high-frequency averaging seems to be dependent on the specific reconstruction method.  For example, at low-SNR, the optimized averaging pattern for SENSE-TV devoted substantially more averages to the center of k-space than the optimized averaging pattern for U-Net reconstruction.  This potentially suggests that the different reconstruction methods are sensitive to slightly different aspects of the data, although the center of k-space appears to be of primary importance in all cases.

\begin{figure}[t]
    \centering
    \includegraphics[width=1\textwidth]{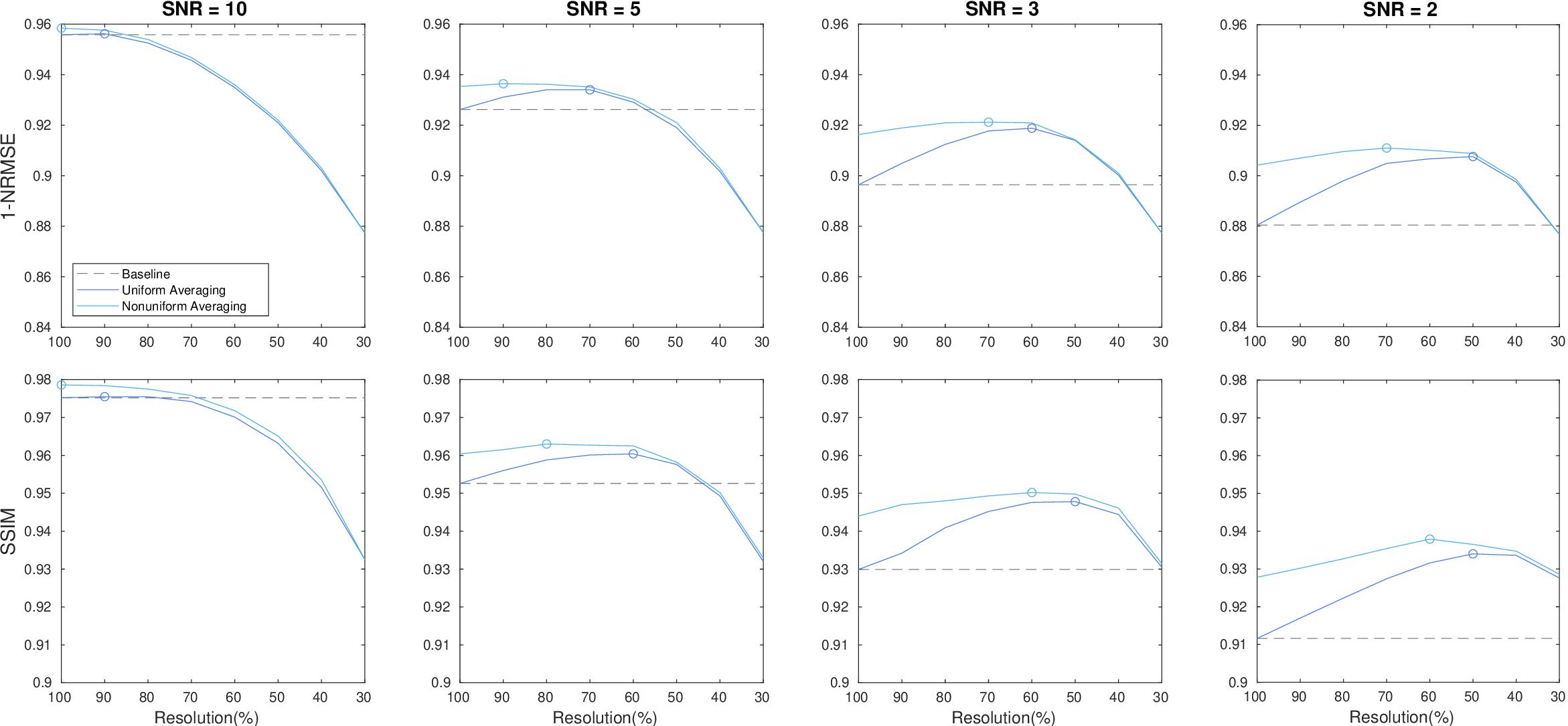}
    \caption{Plots of the quantitative performance of SENSE-TV denoising/reconstruction as a function of resolution, shown for different averaging strategies and different SNR levels.  Each plot shows the mean NRMSE and SSIM values for the uniform averaging and nonuniform averaging schemes for different spatial resolution choices. The best performances for each approach are marked with circles. The resolution of the acquisition is expressed as a percentage relative to the baseline acquisition, i.e., $N/N_0 \times 100\%$.  For reference, we also plot the line corresponding to the results obtained using the conventional high-resolution ($N=N_0$) uniformly-averaged baseline, which matches the first point (100\% resolution) of the uniform averaging curve.  To assist with interpretation, we have chosen to plot 1-NRMSE values (higher is better) in the top row, which is consistent with the SSIM values (where we also have that higher is better) in the bottom row. }
\label{fig:plots}
\end{figure}

\begin{figure}[t]
    \centering
    \includegraphics[width=1\textwidth]{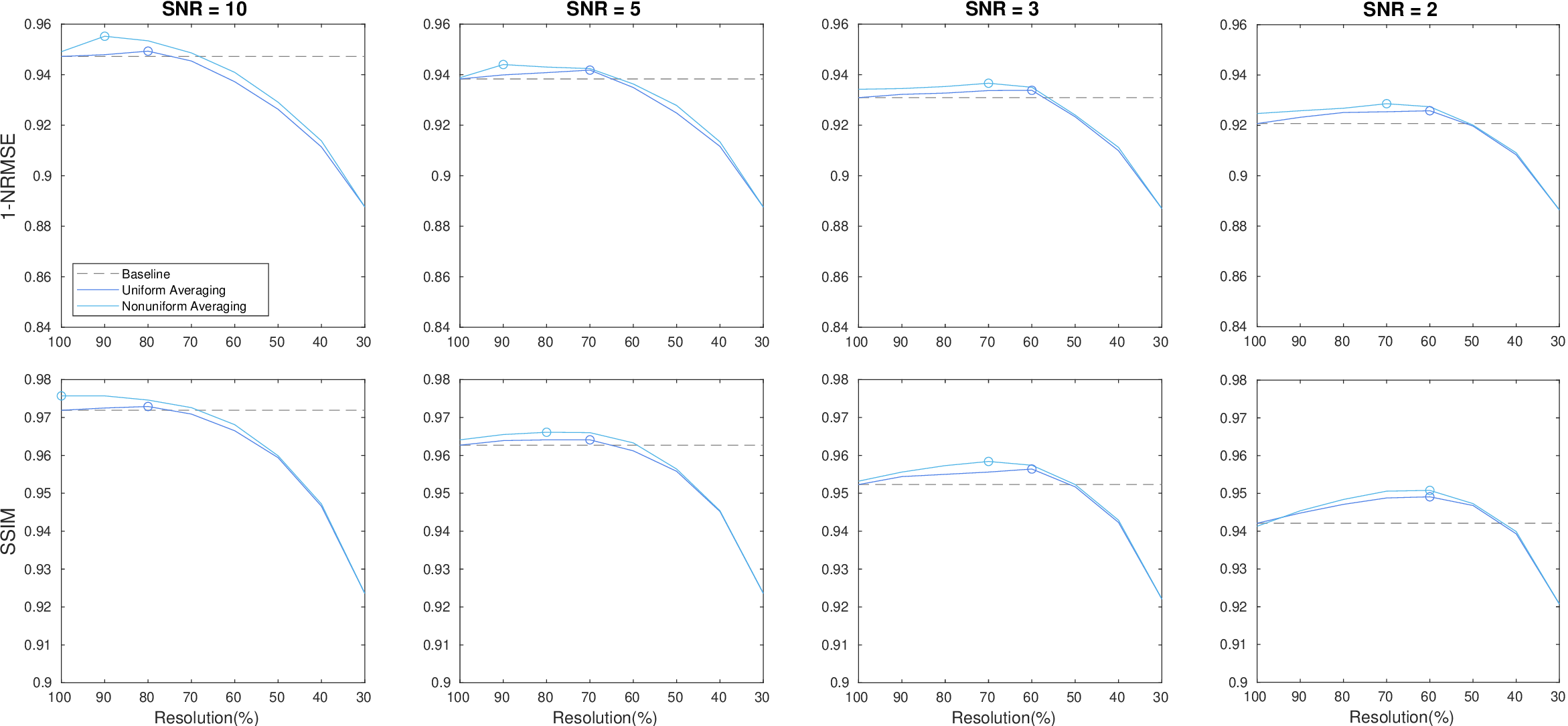}
    \caption{Plots of the quantitative performance of U-Net reconstruction as a function of resolution, shown for different averaging strategies and different SNR levels.  Each plot shows the mean NRMSE and SSIM values for the uniform averaging and nonuniform averaging schemes for different spatial resolution choices.  The best performances for each approach are marked with circles.  The resolution of the acquisition is expressed as a percentage relative to the baseline acquisition, i.e., $N/N_0 \times 100\%$.  For reference, we also plot the line corresponding to the results obtained using the conventional high-resolution ($N=N_0$) uniformly-averaged baseline, which matches the first point (100\% resolution) of the uniform averaging curve.  To assist with interpretation, we have chosen to plot 1-NRMSE values (higher is better) in the top row, which is consistent with the SSIM values (where we also have that higher is better) in the bottom row. }
\label{fig:plotu}
\end{figure}

To gain further insights, Figs.~\ref{fig:plots} and \ref{fig:plotu} show quantitative NRMSE and SSIM performance for SENSE-TV and U-Net reconstruction as a function of spatial resolution.  As can be seen in both cases, using lower-resolution acquisitions ($N\ll N_0$) can offer substantial NRMSE and SSIM benefits compared to using  high-resolution acquisition ($N=N_0$), particularly as the SNR becomes lower.  These plots also show that optimized nonuniform averging achieved slightly better mean NRMSE and SSIM values than were obtained using uniform averaging.  In addition, the optimized gridsizes $\hat{N}$ for nonuniform averaging were often higher than the corresponding optimized $\hat{N}$ obtained for uniform averaging.  

\newpage

\begin{figure}[t]
    \centering
        \includegraphics[width=1\textwidth]{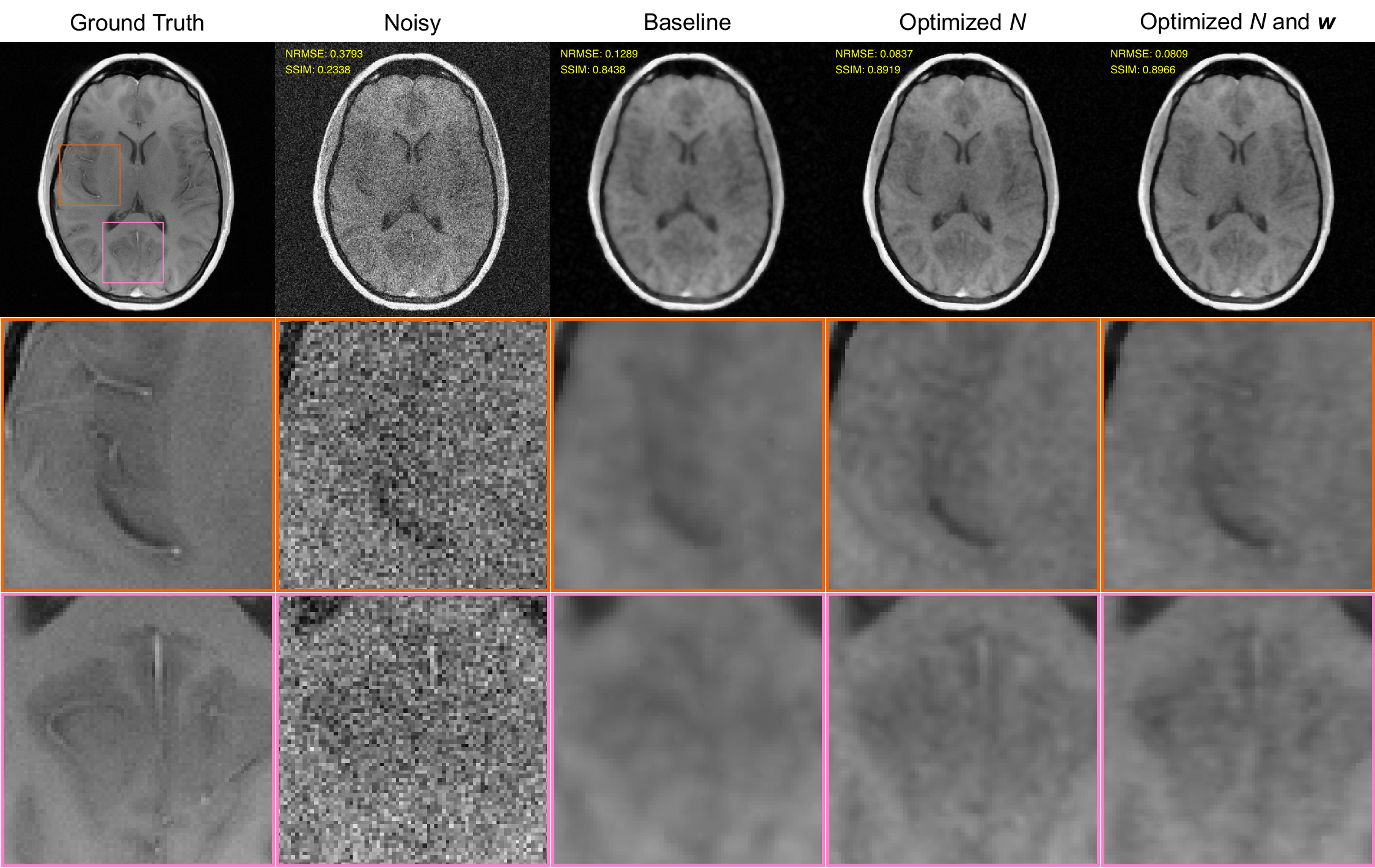}
    \caption{Illustrative results of SENSE-TV denoising/reconstruction for the SNR=2 case, using different acquisition strategies. The left two columns respectively show the ground truth image and the results of simple Fourier reconstruction of the noisy data.  The remaining columns respectively show (from left to right): the denoising/reconstruction results obtained with the baseline high-resolution acquisition with uniform averaging, optimized acquisition with uniform averaging, and optimized acquisition with nonuniform averaging. This figure otherwise uses the same formatting as Fig. 1 (see the Fig. 1 caption for  details). }
\label{fig:sresult}
\end{figure}

\begin{figure}[t]
    \centering
        \includegraphics[width=1\textwidth]{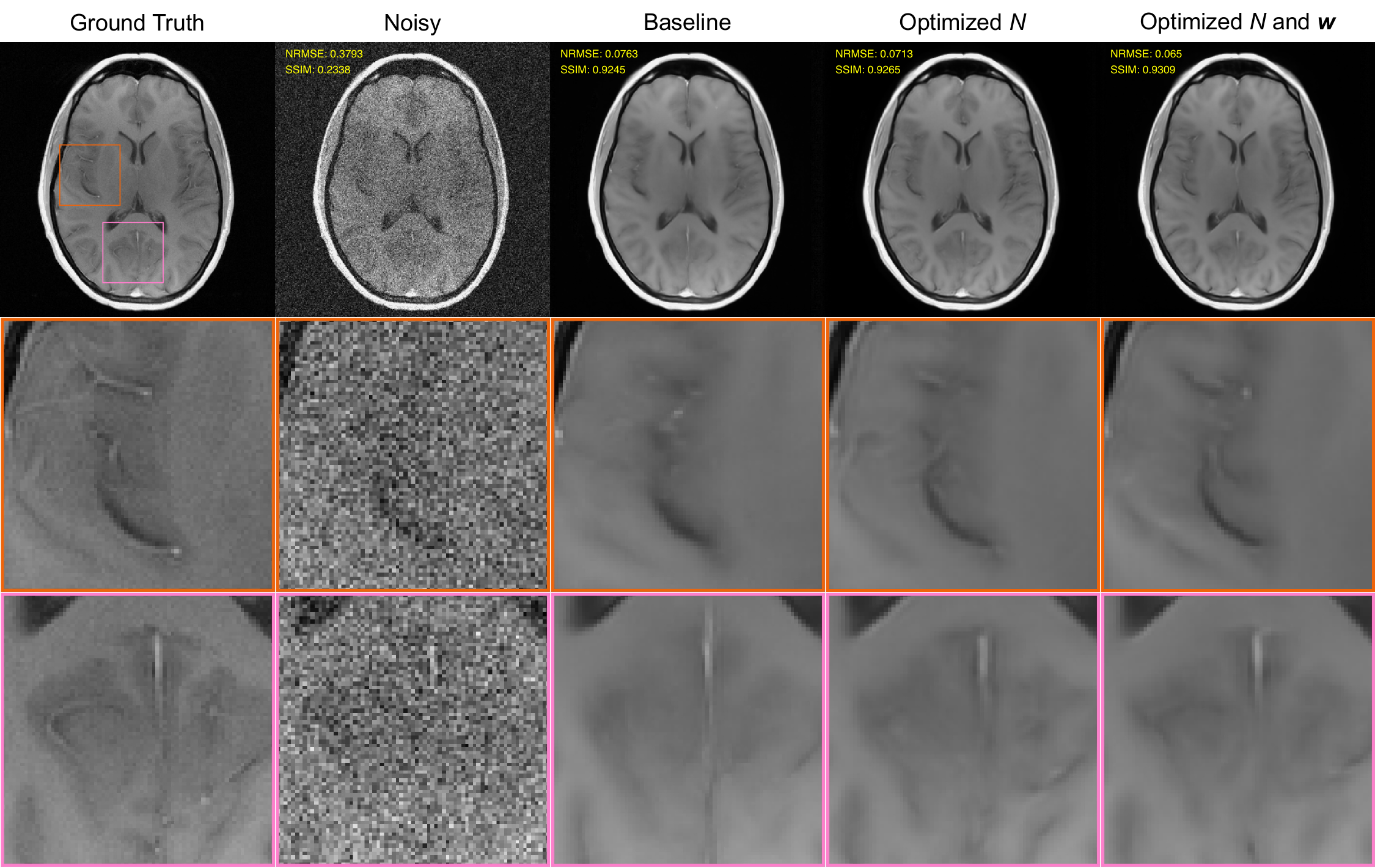}  
\caption{Illustrative results of U-Net denoising/reconstruction for the SNR=2 case, using different acquisition strategies. The left two columns respectively show the ground truth image and the results of simple Fourier reconstruction of the noisy data.  The remaining columns respectively show (from left to right): the denoising/reconstruction results obtained with the baseline high-resolution acquisition with uniform averaging, optimized acquisition with uniform averaging, and optimized acquisition with nonuniform averaging. This figure otherwise uses the same formatting as Fig. 1 (see the Fig. 1 caption for  details). }
\label{fig:uresult}
\end{figure}

The NRMSE and SSIM values shown in our previous plots can be imperfect measures of image quality and can lack sensitivity to important factors such as spatial resolution and hallucination \cite{kim2018a,antun2020,chan2021}. As a result, it is also important to look closely at the denoising/reconstruction results.  Representatitve images are shown for SENSE-TV and U-Net denoising/reconstruction in Figs.~\ref{fig:sresult} and \ref{fig:uresult}, respectively, corresponding to the SNR=2 case. 
 Our subjective assessment of these images suggests that optimizing $N$ indeed seems to yield meaningful visual improvements in image quality over using the baseline high-resolution acquisition with $N=N_0$.     Specifically, while the denoising/reconstructions in this case have visibly lower-resolution than the ground truth images (as should be expected  with $N \ll N_0$), we believe that the resulting images still provide substantially more faithful visual depictions of high-resolution vascular and gray matter structures compared to the images from the high-resolution baseline. On the other hand, while nonuniform averaging produces slightly better NRMSE and SSIM values than uniform averaging, the visual differences between these two approaches are more subtle, and it is difficult to identify a clear winner.  This is a common issue in the field, and speaks to the continuing need for better ways of evaluating image denoising/reconstruction performance \cite{kim2018a,chan2021,wang2024,antun2020}.

\begin{figure}[t!]
    \centering    
        \includegraphics[width=1\textwidth]{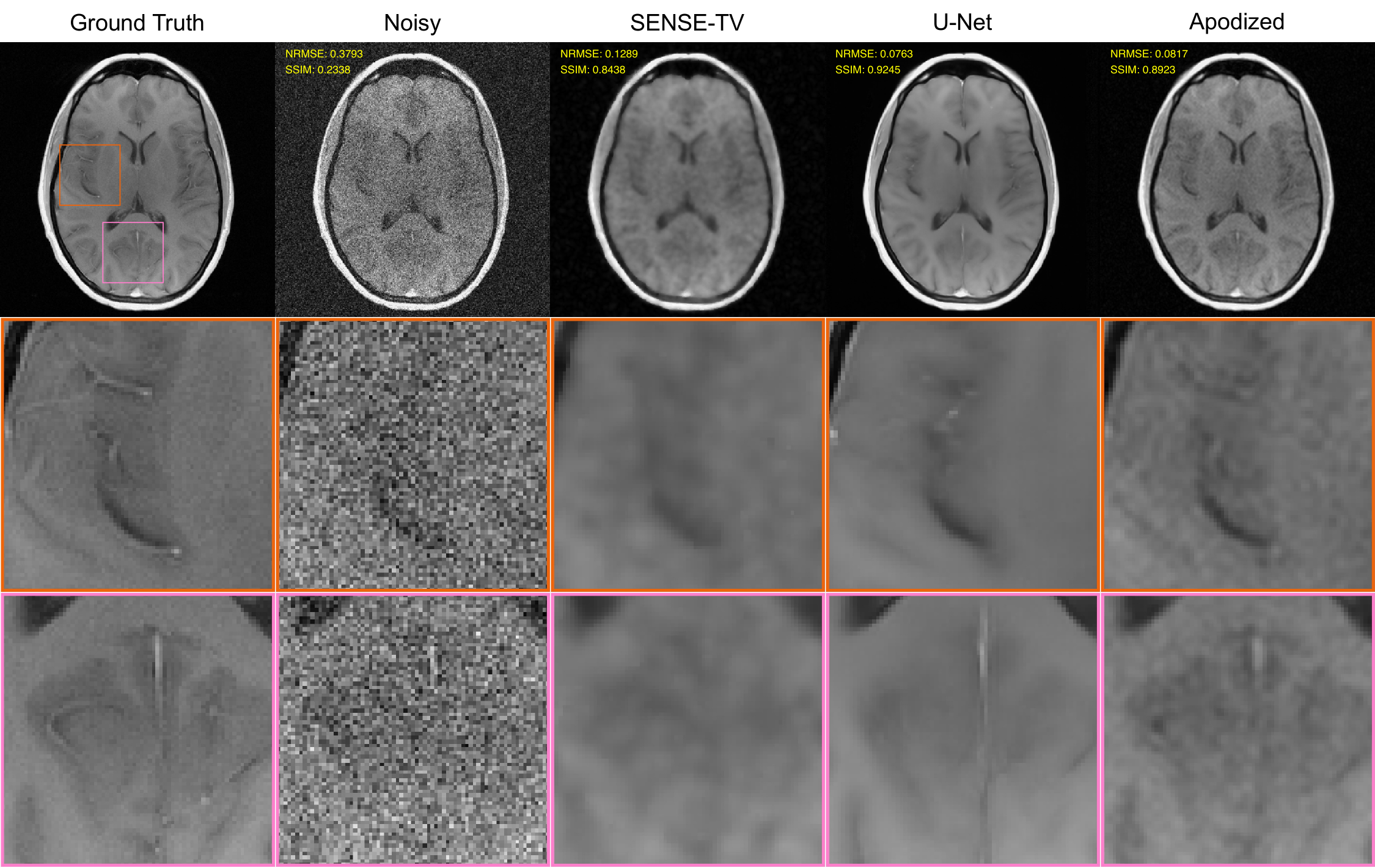}
\caption{A comparison between advanced denoising reconstruction techniques (SENSE-TV/U-Net) with unoptimized acquisition versus simple linear filtering/apodized reconstruction with optimized acquisition.  Results are shown for the SNR=2 case.   The left two columns respectively show the ground truth image and the results of simple Fourier reconstruction of the noisy data.  The remaining columns respectively show (from left to right): the denoising/reconstruction results obtained with SENSE-TV and the baseline high-resolution uniformly-averaged acquisition, U-Net baseline high-resolution  uniformly-averaged acquisition, and apodized reconstruction with optimized nonuniform averaging (i.e., optimized $N$ and $\mathbf{w}$).   This figure otherwise uses the same formatting as Fig. 1 (see the Fig. 1 caption for  details). }
\label{fig:apod}
\end{figure}

Finally, illustrative results comparing simple linear filtering/apodized reconstruction with optimized acquisition ($N$ and $\mathbf{w}$) against  SENSE-TV and U-Net denoising/reconstruction with the baseline acquisition (unoptimized high-resolution acquisition with uniform-averaging) are shown in Fig.~\ref{fig:apod}.  Although the NRMSE and SSIM values in this case suggest that the U-Net offers the best quantitative performance, we personally believe that these metrics are misleading in this case.  Instead, we prefer the simple apodized result, which faithfully reproduces many high-resolution vascular and gray matter features that are more difficult to discern in the U-Net and SENSE-TV results.  This underscores the importance of designing data acquisition carefully, since even very simple denoising methods with good data acquisition designs can yield major visual advantages over over advanced denoising/reconstruction methods with poorly-chosen data acquisition designs.

\section{Discussion}\label{sec5}

In this work, we investigated the extent to which optimized data acquisition strategies could be used to enhance the performance of advanced computational MRI denoising techniques, motivated by insights from the early MRI literature about the tradeoff between SNR and spatial resolution.  Our results suggest that enhanced denoising performance can be achieved by optimizing the k-space coverage of the acquisition.  Indeed, we observe improved quantitative metrics (i.e., improved NRMSE and SSIM) from the use of optimized acquisition design, and we also observe qualitative visual changes in the images that we (subjectively) believe  reflect practical improvements. This is consistent with previous findings \cite{haldar2011,haldar2011c} suggesting that tackling challenging denoising problems may be suboptimal and inefficient, when acquisition modifications could instead be leveraged to make denoising easier.  These results highlight the interplay between acquisition and denoising and the potential advantages of approaching them jointly from an integrated perspective, rather than compartmentalizing denoising away from experimental details.

While  our results were obtained in a specific context (i.e., T1-weighted brain data with 0.5mm in-plane resolution  with a specific set of SNRs),  we expect the core principles to generalize beyond this case.  Indeed, over many years of pursuing denoising research across a wide range of field strengths, spatial resolutions, nuclei, and applications, our group has  never observed a denoising scenario in which the general behavior of thermal noise was distinct from that of any other scenario.  Specific details, such as the optimal acquisition resolution, the optimal averaging pattern, or the amount of performance improvement will of course depend on the application, but our central finding -- that denoising performance can benefit from optimized k-space coverage and optimized k-space averaging -- should be widely applicable.  Indeed, our work is simply building on general theoretical principles from earlier MRI literature that were derived in a context-independent way.  

In addition, while our results were obtained considering SNRs $\leq$ 10, the results can still be relevant to higher-SNR cases.  After all, the way we count averages depends on how we define a single repetition.  Mathematically, there is a negligible distinction between $10\times$ averaging of data with SNR=10 versus $5\times$ averaging of data with SNR=$10\sqrt{2}\approx 14$. These fundamental scaling principles make our insights applicable across a wide range of SNR regimes.

One limitation of our study is that our optimization and evaluation procedure relied on quantitative metrics like NRMSE and SSIM, which are known to be imperfect measures of image quality and can be insensitive to important things like spatial resolution and hallucination \cite{kim2018a,chan2021,wang2024,antun2020}.  Indeed the  lack of sensitivity to spatial resolution is obvious from Figs.~\ref{fig:plots} and \ref{fig:plotu}, where we can still obtain relatively good NRMSE and SSIM values with respect to the original high-resolution reference image, even when using acquisitions with $>50\%$  of the original spatial resolution.  As such, despite the fact that our sampling-optimization approach appears to have produced real visual improvements in image quality, we also believe that any approach that relies solely on metrics such as NRMSE and SSIM to optimize acquisition or reconstruction is likely to be misled to some extent by the undesirable behavior of these metrics.  This can be viewed as a simple manifestation of Goodhart's law, which is popularly stated as ``when a measure becomes a target, it ceases to be a good measure." While Goodhart's law is relatively well-known in many fields, we believe that it is particularly important for the MRI community to have more awareness of the potential limitations of metrics like NRMSE and SSIM, which are increasingly being used to optimize acquisition and reconstruction strategies across a range of different applications.  As our results suggest, it is possible to get relatively good NRMSE and SSIM values, even though the resulting images may not accurately depict image features that may be important for some applications.

Another limitation is that, although our experiments were based on real MRI data and employed realistic models of MRI noise behavior, our simulations also neglected practical experimental nonidealities that would interact with our acquisition manipulations (e.g., effects from eddy currents, field inhomogeneity, and relaxation during the readout).  The choice to neglect these experimental nonidealities was intentional and follows similar practices from the early literature on SNR efficiency in MRI \cite{edelstein1986, parker1990, macovski1996}.  Specifically, this enabled us to perform well-controlled experiments that allow us to study  general  SNR/resolution tradeoff principles in isolation from other factors that are likely to be more sequence- and application-dependent.   This allows us to draw more generalizable conclusions, although also suggests the need for application-specific testing with prospectively acquired data.  We also did not investigate non-Cartesian acquisition strategies, nor did we consider methods that vary the local sampling density of the acquisition instead of performing actual averaging \cite{greiser2003}.  We believe that these topics are beyond the scope of the present paper, although represent important and interesting directions for future research.

In the case of nonuniform averaging, it is worth noting that our optimization procedure does not prohibit some entries of $\mathbf{w}$ from going to zero, thereby allowing for the possibility that an undersampling pattern might emerge.  However, our optimization procedure never produced an averaging pattern that included undersampling in any of the cases we tried.  This outcome is consistent with the conventional wisdom that reconstructing underseampled data frequently results in noise amplification \cite{pruessmann1999,robson2008}, making undersampling potentially unwise in SNR-limited scenarios.    At the same time, our optimization problem is highly nonconvex with multiple local minima, and our optimized averaging patterns only represent local minima -- we have not explored the full range of possible sampling strategies.  Therefore, it remains possible (and becomes more likely as the SNR increases) that better-performing averaging patterns that include some degree of undersampling could exist.  Explorating  a wider range of sampling strategies is beyond the scope of this paper, but remains a potentially interesting topic for future research.

\section*{Acknowledgments}
This work was supported in part by NIH grants R01-MH116173, R01-NS074980, R56-EB034349 and U01-HL167613, the Ming Hsieh Institute for Research on Engineering-Medicine for Cancer, and a USC Annenberg Graduate Fellowship.

\textbf{Data availability:} The datasets used in this study are from the fastMRI database.

\textbf{Author contributions:} Jiayang Wang and Justin P. Haldar designed the study. Jiayang Wang implemented the methods and performed the experiments. Both authors analyzed the results and wrote the manuscript.

\bibliography{sn-bibliography}

\end{document}